\documentclass[lettersize,journal]{IEEEtran}
\usepackage{amsmath,amsfonts}
\usepackage{algorithmic}
\usepackage{array}
\usepackage[caption=false,font=normalsize,labelfont=sf,textfont=sf]{subfig}
\usepackage{textcomp}
\usepackage{stfloats}
\usepackage{url}
\usepackage{verbatim}
\usepackage{graphicx}
\def\BibTeX{{\rm B\kern-.05em{\sc i\kern-.025em b}\kern-.08em
    T\kern-.1667em\lower.7ex\hbox{E}\kern-.125emX}}
\usepackage{balance}
\begin{document}

\title{Rethinking Internet Communication Through LLMs: How Close Are We?}

\author{Sifat Ut Taki and Spyridon Mastorakis
\IEEEcompsocitemizethanks{
    \IEEEcompsocthanksitem{S. U. Taki and S. Mastorakis are with the Department of Computer Science and Engineering, University of Notre Dame, USA (email: staki@nd.edu, mastorakis@nd.edu).}
}
}

\markboth{Journal of \LaTeX\ Class Files,~Vol.~18, No.~9, September~2020}%
{How to Use the IEEEtran \LaTeX \ Templates}

\maketitle

\begin{abstract}

In this paper, we rethink the way that communication among users over the Internet, one of the fundamental outcomes of the Internet evolution, takes place. Instead of users communicating directly over the Internet, we explore an architecture that enables users to communicate with (query) Large Language Models (LLMs) that capture the cognition of users on the other end of the communication channel. We present an architecture to achieve such LLM-based communication and we perform a reality check to assess how close we are today to realizing such a communication architecture from a technical point of view. Finally, we discuss several research challenges and identify interesting directions for future research.

\end{abstract}

\begin{IEEEkeywords}
Internet, Large Language Models, Computer Networks, Communication.
\end{IEEEkeywords}

\section{Introduction}
\label{sec:intro}

Large Language Models (LLMs), such as GPT-3 and BART, represent a significant development in the field of Artificial Intelligence (AI)~\cite{wei2022emergent}. LLMs are typically trained on large volumes of textual data and are able to provide human-like, textual responses to user questions and perform various language-based tasks requested by users. LLMs have been viewed by many as a step for AI to get closer to realizing human cognition. 

Such technological developments were not expected when the Internet first started. The Internet has enabled communication across long distances among individuals in ways that no one had imagined: emails, texting (chatting) applications, voice and video calls. The design of the Internet has not come without concerns though. The research community has raised concerns about cybersecurity attacks, Internet centralization (consolidation), and the ability of the Internet to scale as the volume of network traffic grows year by year~\cite{huitema2023report, zhang2023revealing}.

Keeping in mind the Internet design and the concerns that have been raised about it, as well as the recent advancements of LLMs, in this paper, we envision the use of LLMs that can capture the cognition of individual Internet users. Internet users, instead of communicating directly with each other over the Internet, they communicate with the LLMs of other users. These LLMs can be downloaded by users that initiate communication with others and can be queried locally, or can be deployed remotely (e.g., at the edge of the network or on nearby clouds) in order to be queried by multiple users at the same time. This LLM-based communication vision has the potential to reduce the reliance of users on the Internet as well as the overall volume of generated network traffic. In this paper, our contribution is two-fold:

\begin{itemize}

\item We realize and propose an architecture for LLM-based communication among Internet users. 

\item We conduct a reality check in terms of how close we are today to the technical realization of such a communication architecture and we discuss several open challenges and directions for future research.

\end{itemize}

The rest of this paper is structured as follows. In Section~\ref{sec:overview}, we provide a brief overview of the Internet and discuss how LLMs are relevant when it comes to communication among users over the Internet. In Section~\ref{sec:arch}, we present an architecture and design for LLM-based communication. In Section~\ref{sec:reality}, we conduct a reality check to evaluate how close we are today to realizing this LLM-based communication vision. In Section~\ref{sec:challenges}, we discuss several open challenges and directions for future research. Finally, in Section~\ref{sec:conclusion}, we conclude our paper.

\section{A Brief Overview of the Internet and How LLMs Are Relevant}
\label{sec:overview}

The Internet started as a military project of the Advanced Research Project Agency (ARPA) with the goal of enabling robust communication during cold war if a nuclear attack against the United States was launched~\cite{lukasik2010arpanet}. In the 1980s, connections and links to US universities were added. With the inception of the Web in the 1990s, the Internet became commercial. Since the 1990s, the Internet has widely enabled communication among humans in various ways and through several applications, such as messaging/chatting, as well as voice and video calling. Over the years, several challenges have been highlighted and several concerns have been raised about the operation of the Internet, including various types of cyberattacks (e.g., Distributed Denial of Service (DDoS) attacks, attacks against the Domain Name System, and security breaches)~\cite{zhang2023revealing}, concerns about the scalability of the Internet architecture as the volume of network traffic grows year by year, as well as concerns about the consolidation (centralization) of the Internet~\cite{huitema2023report}.

With the recent proliferation of LLMs, and predominantly ChatGPT, humans have been able to essentially interact (in human readable and understandable context) with these LLMs. For example, users can type questions, which are parsed by LLMs, and LLMs type responses back to users. Keeping the Internet design, the concerns raised about it, and the recent LLM advancements in mind, in this paper, we envision the use of LLMs to capture the cognition of individual Internet users. As a result, instead of communicating with other users over the Internet, users can communicate with the LLM of others, which can be deployed either locally or remotely (at the edge or on a nearby cloud). This reduces the reliance of users on the Internet operation as well as the overall volume of generated Internet traffic, since users can communicate with LLMs of other users deployed locally or physically close to them.
For example, in Figure~\ref{fig:arch}, user A would like to communicate with user C. User C provides data for the creation and training of their LLM, while user A can download user C's LLM and query it locally or user C's LLM can be deployed on edge/cloud nodes in order to be queried by multiple users at the same time (e.g., users A and B in Figure~\ref{fig:arch}).

\section{Architecture}
\label{sec:arch}

In this section, we present the main design and the architecture that leverages LLMs for communication between users. The following aspects are discussed in this section: creating and training an LLM with user data to capture the cognition of an individual user, deploying/offloading trained LLMs close to other users, and retraining LLMs with up-to-date information from users to capture new new life updates. Finally, we present a workflow to showcase the overall operation of the architecture.

\subsection{Training personal LLMs}

The first part of the process is to train individual LLMs with data provided from each user, so that an LLM can learn information about a user. In other words, each user will have their own (personal) LLM. Users can train their LLMs either locally (if they have access to appropriate hardware) or the training process can happen in a data center. As a training dataset, data from previous conversations of the user can be used, or the user can select other data to be used for training. Once an LLM learns about a user with an acceptable accuracy, the LLM then can be used for natural communication with other users. LLMs can be trained from scratch, or pre-trained LLMs can be used to reduce the overall training time.

\subsection{Deploying/offloading LLMs}
Once LLMs are trained with sufficient data and an acceptable accuracy, The inference-only version of the model can be offloaded to edge or cloud nodes closer to other users. Alternatively, users could download LLMs of the users they would like to communicate with locally if they have adequate hardware resources. In these cases, users can either run prompts on the edge/cloud nodes or run prompts locally if they have downloaded LLMs. 

\begin{figure}
    \centering
    \includegraphics[width=1\linewidth]{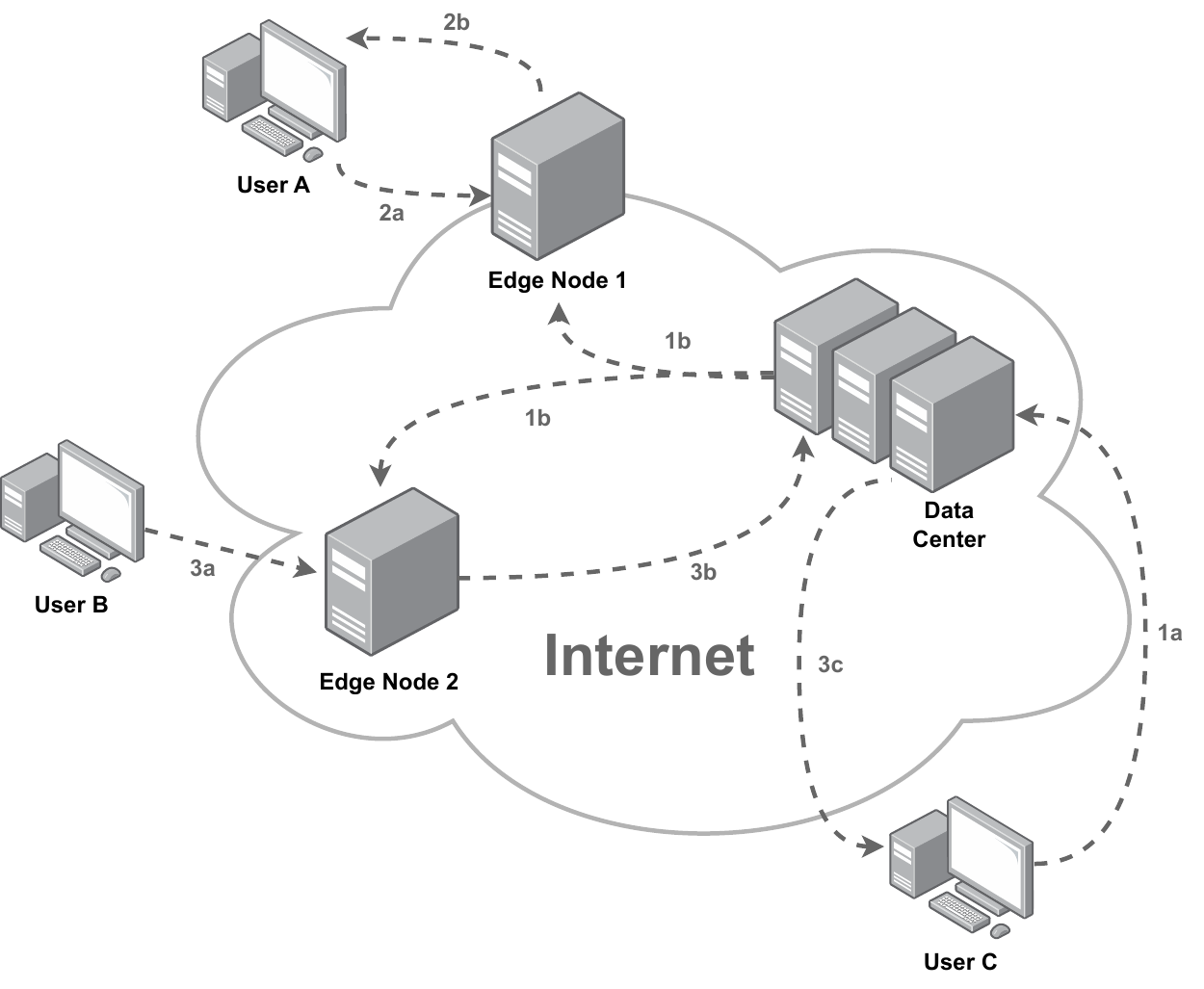}
    \caption{Architecture of an LLM-based communication system.}
    \label{fig:arch}
\end{figure}

\subsection{Updating trained LLMs}

Trained LLMs may need to be updated over time with up-to-date information about users. The models can also learn from the actual responses of users, when users participate in a conversation. Once the models are updated with new data/information, the new models will be offloaded to edge/cloud nodes or will be downloaded by other users, thus replacing the old models. 

\subsection{Workflow}

\begin{figure}
    \centering
    \includegraphics[width=1\linewidth]{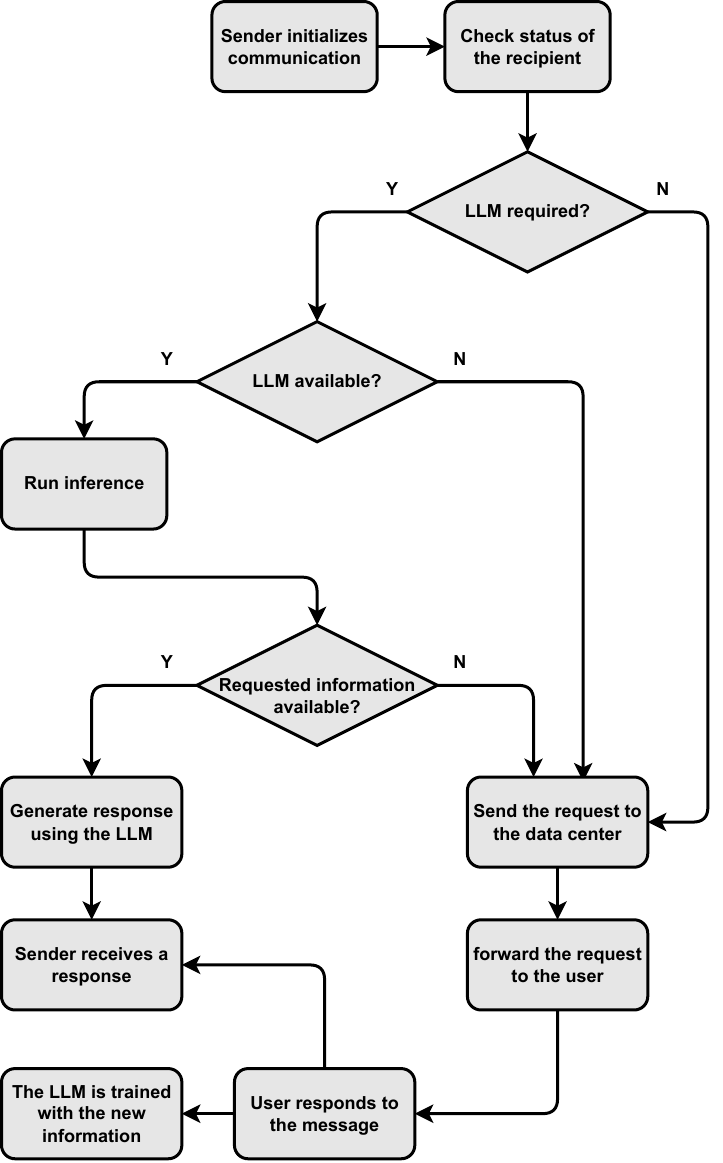}
    \caption{Workflow of the proposed LLM-based communication system.}
    \label{fig:workflow}
\end{figure}

The workflow of our architecture is presented in Figure~\ref{fig:workflow}. The way that our architecture will handle a request from a user to communicate with another user depends on the current status of the recipient. A user has the following options for their current status, which are discussed below:

\noindent\textbf{\textit{Active:}} One option for the user is to be \textit{active}. This status is like any other currently available chat system where every message will be forwarded to the inbox of the original recipient, and the original recipient will be responsible for the response. However, the recipient will still have the option to generate a response using an LLM if they want to.

\noindent\textbf{\textit{Busy:}} In this status, the recipient will only receive messages directly to the inbox if they are from a certain group of senders, who the recipient has selected beforehand. All the other messages will be served by the LLM for that particular recipient (provided that there is an LLM available). Responses generated by the LLMs will contain a note (e.g., \textit{"This is an AI-generated message"}), which will help the original sender know that responses are AI-generated and not from the original user. This status is helpful when a user is in a situation where they only want to directly respond to their close friends and family. The servers hosting LLMs will keep logs of the original queries and the responses generated by LLMs for users in \textit{busy} status to review later.

\noindent\textbf{\textit{Away:}} Another status for the user is to be \textit{away}. In this status, every message will be responded by the LLM (provided that an LLM available). This will be helpful in situations where a user is not able to respond to messages and would like to completely rely on the LLM. Again, all responses generated by LLMs will contain a note to let senders know that these are AI-generated messages, and the servers hosting the LLMs will keep logs of queries and LLM-generated responses. In both \textit{Busy} and \textit{Away} states, there could be cases where the model is unable to generate a response because of lack of information. In those cases, the messages will be forwarded to the recipient and the recipient will have the option to respond later when they are \textit{active}.

\noindent\textbf{\textit{Inactive:}} Finally, a user can be in the \textit{inactive} state. In this state, neither the recipient will receive the message, nor the LLM will generate a response. The message will be held in the inbox until the user becomes \textit{Active}; after which, the user can either respond himself, or let the LLM generate a response for him.


Figure~\ref{fig:arch} presents an example of communication among users based on our architecture. In this example, User C first provides the necessary data to a data center in order to train a personal LLM (Step 1a). Once the model is ready, it is then offloaded to edge nodes close to User A and User B (Step 1b). When User A sends a message to User C, the device of User A and edge node 1 checks the current status of User C and whether the LLM is required or not (Step 2a). In our example, the LLM will generate a response on behalf of User C, which is available on edge node 1. As such, edge node 1 will run an inference to check if the question is answerable by the model. The model will generate an answer and will send it to User A (Step 2b). Subsequently, User B will try to communicate with User C. The message from User C will be processed by edge node 2 (Step 3a). However, the involvement of User C will be required to generate a response in this case. As a result, the message will be forwarded to the data center (Step 3b). The data center will then forward the message to User C (Step 3c). When User C responds back to User B, the LLM will learn from the new information.

\subsection {Extending Beyond Textual-Based Communication}

We envision our architecture to extend beyond text-based (messaging) communications. Text-to-speech is a well-established technology that can transform text into human speech~\cite{kaur2023conventional}. With the help of modern AI, it is now possible to synthesize the voice of a specific person. A research team at Microsoft has developed a neural codec language model called VALL-E~\cite{wang2023neural}. VALL-E is capable of leveraging in-context learning to synthesize high-quality personalized audio from a recording of only 3 seconds. VITS-2~\cite{vits2} is another AI-based speech synthesizer that was proposed recently with considerably better performance. In the future, we expect text-to-speech technologies to become effective to a point where it will be impossible to discern whether the voice is AI-generated or from an actual person. In the context of our architecture, the responses generated by personalized LLMs can be converted to audio speech. As a result, our architecture will be able to offer voice communication as well.

The sample applies to offering video-based communication through our architecture. Deepfake-based video generations are getting exceptionally good at mimicking the appearance of a person in pre-recorded videos~\cite{westerlund2019emergence}. Using deepfake-based solutions, such as DeepFaceLab~\cite{deepfacelab}, videos of a person talking can also be generated. Even with the technology we have today, it is difficult to discern between real videos and videos generated by Deepfake AI. 
By combining the responses generated by the LLM, voice generated by advanced AI-based text-to-speech approaches, and videos of the recipient user generated by advanced Deepfake AI, our architecture will be capable of offering the full experience of video-based communication produced entirely by AI.


\section{Reality Check: How Close Are We?}
\label{sec:reality}

Commercial implementations of chat bots are very popular nowadays. A chat bot can handle basic queries from a customer and generate informative responses without any involvement from an actual person, thus reducing the overall operating costs of a business. However, a simple chat bot has rather limited capabilities. Another relevant service is \textit{Character.Ai}~\cite{CharAI}, which allows users to either chat with AI-generated characters trained on personal likeness of famous people, or create their own character with their own data. This service is also limited as the machine learning models used are not capable enough to capture human cognition. With the emergence of LLMs, more capable and sophisticated messaging systems can now be implemented.

LLMs have seen a boom in recent years. ChatGPT by OpenAI can generate human-like responses from a natural prompt by a user. Given enough data, an LLM can learn contextual information and generate responses accordingly. Although researchers have yet to design and train an LLM to specifically mimick and represent a real person, Meta has already started to build an AI that can process language like a real human being~\cite{meta-ai}. They have found similarities between brains and LLMs from their initial research. As a result, we foresee the possibility to develop LLMs personalized with the data of an individual user in the future, which should be able to generate responses representing that user.

Current LLMs are extremely large with billions of trainable parameters. This is a challenge for LLM-based communication approaches. Training such large LLMs require massive computation power. However, researchers have been actively working on reducing the number of parameters required to achieve acceptable performance. Reducing the number of parameters will allow an LLM to train with considerably fewer FLOPs, or with more tokens. Recently announced Llama-2 by Meta AI with only 7 billion parameters can achieve similar performance to MPT with 30 billion parameters and the 70 billion version of the Llama-2 model can be as good as GPT-3.5 with 175 billion parameters~\cite{llama-2}. As the research in this field progresses, we expect to see more efficient models being proposed in the future. Moreover, advancements in computer hardware are also expected to help with training LLMs for individuals efficiently.

\begin{figure}
    \centering
    \includegraphics[width=\linewidth]{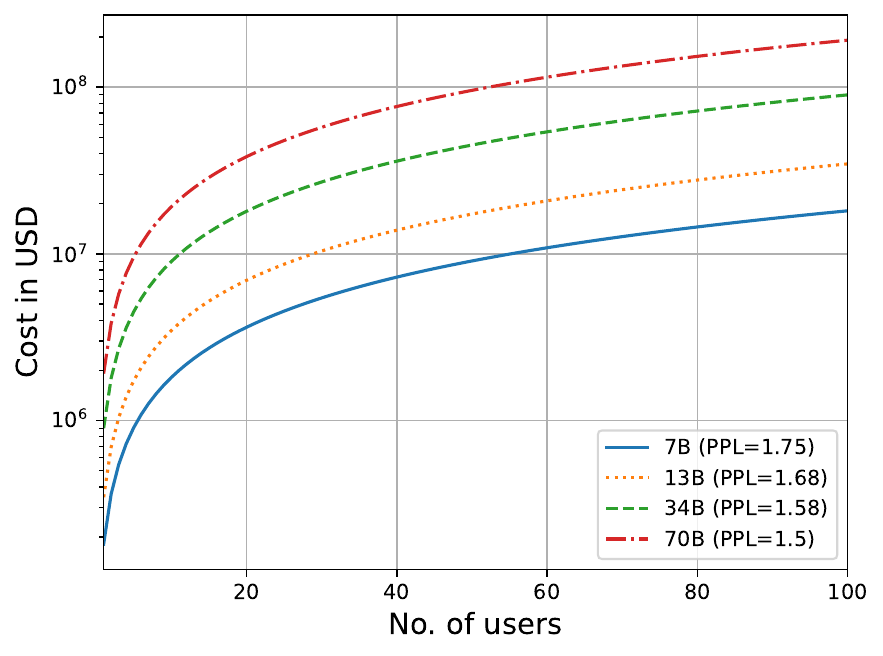}
    \caption{Cost of training different Llama2 models with the lowest corresponding PPL.}
    \label{fig:llama_cost}
\end{figure}

Training an individual LLM for each user is a major challenge when it comes to the substantial computational resource requirements. Figure~\ref{fig:llama_cost} presents the cost in USD for training different versions of Llama models. The cost was estimated for the NVIDIA A100 GPU. As we can see from the figure, training a single Llama2 model with 7 billion parameters can cost more than 180 thousand USD, whereas the largest Llama2 model with 70 billion parameters can cost up to 2 million USD to reach a Perplexity (PPL) of 1.5~\cite{llama-2}. This number quickly adds up when we need to train an individual LLM for each user. Moreover, the carbon footprint caused by training the models will also be substantial. However, we can leverage pre-trained models, so that there is no need to train every single model from scratch for multiple users. 

To further assess the feasibility of an LLM-based communication architecture based on currently available technologies, we have implemented a small-scale prototype with two modules: an LLM-based response generator and an AI-based text-to-speech converter. The response generator was built using the 7B version of the Llama2 model to generate responses and the VITS2 text-to-speech audio synthesizer was used to convert the responses to personalized voice responses. The prototype was evaluated on a system with an AMD Ryzen Threadripper PRO 5955WX (16-Cores), 128 GB of system memory, and an NVIDIA GeForce RTX 4090 with 24 GB of video memory. Table~\ref{tab:res_time} presents the average load time and the process time of different components when generating responses with 114 tokens and 534 characters. Our results demonstrate that it takes 16.7 seconds to generate a textual response and convert it to a voice response. 

Although the community has made significant advancements in NLP with different LLMs in recent years, we are still not in a position to successfully implement and deploy an LLM-based communication system at a large scale. However, as the technology progresses and we develop more efficient models and hardware, we expect that such a communication architecture will be possible.


\begin{table*}[]
\label{tab:res_time}
\centering
\caption{Response time of different modules for an LLM-based communication architecture.}
\begin{tabular}{|lllll|l|}
\hline
\multicolumn{1}{|l|}{Module}             & \multicolumn{1}{l|}{Model}           & \multicolumn{1}{l|}{Size}    & \multicolumn{1}{l|}{Load Time (s)} & Process Time (s) & Total Time (s) \\ \hline
\multicolumn{1}{|l|}{Response generator} & \multicolumn{1}{l|}{Llama-2-7B-Chat} & \multicolumn{1}{l|}{13.5 GB} & \multicolumn{1}{l|}{6.62}          & 9.64             & 16.26          \\ \hline
\multicolumn{1}{|l|}{Text-to-speech}     & \multicolumn{1}{l|}{VITS2}      & \multicolumn{1}{l|}{15.6 MB} & \multicolumn{1}{l|}{0.26}          & 0.18             & 0.44           \\ \hline
\multicolumn{5}{|l|}{Total Time}                                                                                                                                       & 16.7           \\ \hline
\end{tabular}
\end{table*}

\section{Open Challenges and Directions for Future Research}
\label{sec:challenges}

As of now, there are numerous challenges that need to be tackled for the realization of LLM-based communication. Most of these challenges are due to current limitations in available computing resources and the size of the LLMs. In this section, we discuss the challenges that need to be tackled and we identify directions for future research. 

\noindent\textbf{The shier size of LLMs:} LLMs are extremely large neural network models by nature. An LLM can have anywhere from a few millions to a few billions of trainable parameters. Storing such large models is a critical challenge. The GPT-3 model consists of 175 billion parameter. Even if we use the smallest Llama2 model available now with only 7 billion parameters, it would still require 13.5 GB of free space to store a model. Considering each user will have their own model, the size of the required storage itself will be a massive challenge. Moreover, the datasets required to train the models are massive in size ranging from a few hundred gigabytes to more than a terabyte of data. Considering current technologies, it may be hard to provide the required storage capacity. To overcome this challenge, further research in needed in terms of reducing the size of LLMs and increasing the capacity of edge/cloud data centers to accommodate the storage of additional LLMs.


\noindent\textbf{Effective training methods of LLMs:} As discussed earlier, training LLMs is another challenge because of their sizes as well as the capability of the hardware we currently have. However, as LLMs become smaller and more capable, and the available hardware capabilities advance, it will become possible to train LLMs with more tokens. Moreover, instead of training every model from scratch, transfer learning techniques can be leveraged to drastically reduce the time and resources required to train LLMs. Finally, mechanisms for the reuse of computation can be leveraged to speed up the training process of LLMs and reduce the usage of computing resources~\cite{al2022promise}.

\noindent\textbf{Energy consumption and carbon footprint:} Even if we have enough hardware resources to store, train, and run inference on the LLMs, the energy requirements for running such a system will be tremendous. When trained with a cluster of NVIDIA A100 GPUs, the largest Llama2 model with 70 billion parameters consumes 688,128 kWh of energy, emitting 291.42 tCO2eq of carbon. The smaller 7 billion parameters version consumes 73,728 kWh of energy and emits 31.22 tCO2eq of carbon~\cite{llama-2}. These numbers increase with the number of individuals that use our LLM-based architecture for communication. To resolve this challenge, LLM optimizations and transfer learning approaches can be explored. Meanwhile, more efficient hardware like the recently announced Nvidia Grace Hopper superchips with The NVLink Chip-2-Chip interconnect between the CPU and GPU, which provides 7x more bandwidth and 4.5x higher throughput while consuming more than 5x less energy than the currently available top-tier NVIDIA H100 GPU~\cite{nvidia_gracehopper}.



\noindent\textbf{Accuracy of the models:} The LLMs are very good at generating human-like responses; however, these models are never 100\% accurate with the information that they provide. The responses from the LLMs can be inaccurate form time to time, and there is no way for the receiver to tell whether the response is correct or not. Advancements in LLM fine-tuning will help achieve acceptable accuracy. Moreover, developments in prompt engineering and safety checkers can help prevent LLMs from generating responses that contain false information.

\noindent\textbf{Real-time responses:} In our architecture, there will be times that real-time responses will be required from the original recipient. Each model will have safety checker mechanisms in cases that a user would like to learn information that the model is not currently aware of. Mechanisms to build such safety checkers that work effectively need to be explored.


\noindent\textbf{Security:} Security is a critical aspect when realization such an LLM-based communication architecture. Service providers need access to sensitive information of each user in order to properly train each LLM. To tackle this challenge, privacy-preserving mechanisms for LLM training need to be explored.
Moreover, users can choose what information they want share with each other during real-time communication; however, a user can get any information from the LLM regardless of the choice of the recipient. To this end, mechanisms to allow users to select what information their LLMs can expose to communicating users and what information to keep private or to share only with a certain group of communicating users need to be developed.


\section{Conclusion}
\label{sec:conclusion}

In this paper, we presented our vision for the realization of an LLM-based communication architecture. We also conducted a reality check to assess how close we are today in terms of technically deploying such a communication architecture. Our evaluation demonstrated that from a technical point of view there are still several challenges that we need to overcome in order to realize such an architecture on a large scale. These challenges are primarily related to the reducing the sheer size of LLMs, advancing LLM storage and training technologies, building energy-efficient frameworks for model training, improving the accuracy of LLMs, and safeguarding the overall operation of our LLM-based communication architecture. We hope that this paper will act as a starting point to motivate the AI, computer systems and networks, and security research communities to explore LLM-based architectures for efficient communication among users. 


\bibliographystyle{unsrt}
\bibliography{sections/ref}

\begin{IEEEbiographynophoto}{Sifat Ut Taki}
is a Ph.D. student at the University of Notre Dame. He received his B.S. in Computer Science and Engineering from Brac University, Bangladesh in 2020. His research interest is in cloud service optimization and privacy.
\end{IEEEbiographynophoto}

\begin{IEEEbiographynophoto}{Spyridon Mastorakis}
is an Assistant Professor in Computer Science and Engineering at the University of Notre Dame, USA. He received his Ph.D. in Computer Science from the University of California, Los Angeles in 2019. His research interests include network systems and architectures, edge computing, and security.
\end{IEEEbiographynophoto}

\end{document}